\PassOptionsToPackage{unicode}{hyperref}
\PassOptionsToPackage{hyphens}{url}
\documentclass[
]{article}
\usepackage{xcolor}
\usepackage[margin=1in]{geometry}
\usepackage{amsmath,amssymb}
\usepackage{iftex}
\ifPDFTeX
  \usepackage[T1]{fontenc}
  \usepackage[utf8]{inputenc}
  \usepackage{textcomp} 
\else 
  \usepackage{unicode-math} 
  \defaultfontfeatures{Scale=MatchLowercase}
  \defaultfontfeatures[\rmfamily]{Ligatures=TeX,Scale=1}
\fi
\usepackage{lmodern}
\ifPDFTeX\else
\fi
\IfFileExists{upquote.sty}{\usepackage{upquote}}{}
\IfFileExists{microtype.sty}{
  \usepackage[]{microtype}
  \UseMicrotypeSet[protrusion]{basicmath} 
}{}
\makeatletter
\@ifundefined{KOMAClassName}{
  \IfFileExists{parskip.sty}{%
    \usepackage{parskip}
  }{
    \setlength{\parindent}{0pt}
    \setlength{\parskip}{6pt plus 2pt minus 1pt}}
}{
  \KOMAoptions{parskip=half}}
\makeatother
\usepackage{graphicx}
\makeatletter
\newsavebox\pandoc@box
\newcommand*\pandocbounded[1]{
  \sbox\pandoc@box{#1}%
  \Gscale@div\@tempa{\textheight}{\dimexpr\ht\pandoc@box+\dp\pandoc@box\relax}%
  \Gscale@div\@tempb{\linewidth}{\wd\pandoc@box}%
  \ifdim\@tempb\p@<\@tempa\p@\let\@tempa\@tempb\fi
  \ifdim\@tempa\p@<\p@\scalebox{\@tempa}{\usebox\pandoc@box}%
  \else\usebox{\pandoc@box}%
  \fi%
}
\def\fps@figure{htbp}
\makeatother
\NewDocumentCommand\citeproctext{}{}

\makeatletter
 \let\@cite@ofmt\@firstofone
 \def\@biblabel#1{}
 \def\@cite#1#2{{#1\if@tempswa , #2\fi}}
\makeatother
\newlength{\cslhangindent}
\newlength{\csllabelwidth}
\newenvironment{CSLReferences}[2] 
 {\begin{list}{}{%
  \setlength{\itemindent}{0pt}
  \setlength{\leftmargin}{0pt}
  \setlength{\parsep}{0pt}
  \ifodd #1
   \setlength{\leftmargin}{\cslhangindent}
   \setlength{\itemindent}{-1\cslhangindent}
  \fi
  \setlength{\itemsep}{#2\baselineskip}}}
 {\end{list}}
\usepackage{calc}

\newcommand{\CSLLeftMargin}[1]{\parbox[t]{\csllabelwidth}{\strut#1\strut}}
\newcommand{\CSLRightInline}[1]{\parbox[t]{\linewidth - \csllabelwidth}{\strut#1\strut}}

\usepackage{booktabs}
\usepackage{longtable}
\usepackage{array}
\usepackage{float}
\usepackage{makecell}
\usepackage{booktabs}
\usepackage{longtable}
\usepackage{array}
\usepackage{multirow}
\usepackage{wrapfig}
\usepackage{float}
\usepackage{colortbl}
\usepackage{pdflscape}
\usepackage{tabu}
\usepackage{threeparttable}
\usepackage{threeparttablex}
\usepackage[normalem]{ulem}
\usepackage{makecell}
\usepackage{xcolor}
\usepackage{bookmark}
\IfFileExists{xurl.sty}{\usepackage{xurl}}{} 
\makeatletter
\@ifundefined{xmpquote}{\newcommand{\xmpquote}[1]{#1}}{}
\makeatother
\hypersetup{
  pdftitle={When bad adjustment looks good: what goes wrong in Plasmode 0.1.0 simulations},
  pdfkeywords={\xmpquote{plasmode simulation; confounding; propensity
score; reproducibility; simulation validity}},
  hidelinks,
  pdfcreator={LaTeX via pandoc}}

\title{When bad adjustment looks good: what goes wrong in Plasmode 0.1.0
simulations}
\author{M. Ehsan Karim\\
School of Population and Public Health, University of British Columbia,
Vancouver, Canada\\
\href{mailto:ehsan.karim@ubc.ca}{\nolinkurl{ehsan.karim@ubc.ca}}\\
ORCID 0000-0002-0346-2871}
\date{}

\begin{document}
\maketitle
\begin{abstract}
Comparisons of confounding-control methods are informative only when
simulated data contain the treatment--covariate relationship those
methods address. We assessed whether \texttt{Plasmode} 0.1.0 preserves
that relationship, uses the returned exposure to generate outcomes, and
reports the effect implied by its generating model. We inspected the
source and ran the generator under identical seeds with source rows as
supplied, sorted by observed exposure, or using an alignment correction.
We ran 200 null-effect replicates per strategy in a synthetic cohort and
the SUPPORT/Right Heart Catheterisation cohort. A sampled subject
usually received another subject's exposure probability. In the
synthetic cohort, source- model AUC was 0.503 before and 0.841 after
alignment. Crude risk-difference bias moved from +0.0051 to +0.3037.
Other orders preserved or reversed the association. After alignment,
deliberately misspecified estimators retained risk-difference biases of
0.10--0.11. Correctly specified estimators were nearly unbiased. With
the unmodified generator, all five appeared similarly unbiased. At an
injected odds ratio of 2, the conditional log odds ratio appeared on the
observed exposure (+0.696; target 0.693), not the returned exposure
(+0.017). Misalignment can weaken, preserve, or reverse designed
confounding and thereby distort method comparisons. Alignment correction
does not repair outcome--exposure decoupling, disagreement between the
reported and generating effects, or marginal miscalibration.
\end{abstract}

\section{Introduction}\label{introduction}

Methods for controlling confounding are routinely compared in
simulation, where the true effect is known and each method's bias can be
measured against it\textsuperscript{1}. Such a comparison is only
informative if the simulated data contain the treatment--covariate
relationship those methods are meant to address. Where the simulated
treatment bears little relation to the covariates, a crude estimate is
already close to the truth, and even a badly specified adjustment can
appear to perform well. Apparent performance then reflects the
simulation rather than the method being evaluated\textsuperscript{2}.

Plasmode simulation answers a different concern: keeping the data
realistic. Instead of drawing covariates from a specified distribution,
it resamples them from a real cohort, so the joint structure of the
covariates --- correlations, skewness and sparsity that would be awkward
to specify by hand --- is whatever the source data happened to
contain\textsuperscript{3}. What is simulated on top of those covariates
varies, and the choice matters. Shaw et al.~distinguish
\textbf{sample-treatment} from \textbf{generate-treatment} designs: the
first resamples each subject's observed treatment along with their
covariates, while the second draws a new treatment from a probability
computed from those covariates. They recommend the latter for evaluating
methods that rely on the propensity score to estimate a point treatment
effect\textsuperscript{2}. That design carries a required sequence:
sample a subject's covariates, draw their treatment from a probability
computed from those covariates, and generate their outcome from the
treatment just drawn\textsuperscript{2,4}.

Any generate-treatment simulation must preserve three correspondences.
The probability used to draw a subject's treatment must be computed from
that subject's own covariates (row correspondence). The outcome must be
generated from the treatment returned for analysis (intervention
correspondence). The reported effect must be implied by the
outcome-generating model (target correspondence).

The \texttt{Plasmode} 0.1.0 R package was intended to support such
simulations, including formula-based exposure
generation\textsuperscript{5}. CRAN archived it in
2021\textsuperscript{6}. It has nevertheless been used in published
methodological work since\textsuperscript{7--9}, although we did not
determine whether those studies used the affected branch.
Generate-treatment plasmode simulation also remains an active approach.
More recent tooling for evaluating estimators within user-supplied data
uses the same design\textsuperscript{10}.

The practical importance of an implementation defect depends on both
what the code does and how often the affected implementation is used. If
the software assigns treatment or generates outcomes by a mechanism
other than the one the analyst specified, the benchmark no longer has
the intended data-generating structure. That discrepancy may not be
apparent from the resulting data.

We examined whether \texttt{Plasmode} 0.1.0 preserves these
correspondences and whether departures from them can alter comparisons
of confounding-control methods.

\section{Methods}\label{methods}

\subsection{Source inspection and alignment
strategies}\label{source-inspection-and-alignment-strategies}

We examined the branch of \texttt{PlasmodeBin()} in which both the
exposure and the outcome model are supplied as formulas, the
configuration intended to produce a generate-treatment simulation. The
package documentation states that when both models are supplied the
simulated outcome depends on the simulated exposure\textsuperscript{5};
that contract is what we tested the execution against. Working from the
archived source, we traced the ordering of the source data frame, the
indices drawn by the sampler, the design matrix and probabilities used
to draw exposure, the design matrix used to generate outcomes, and the
contrast the function returns as the true effect. Source line numbers
and an inventory of the function's remaining branches are in Web
Appendix G.

To test whether a sampled subject receives their own exposure
probability, we compared three ways of running the same simulation under
identical arguments and random-number seeds. The first ran the function
\emph{unmodified} with the source rows in the order supplied (strategy
A). The second ran it \emph{unmodified after sorting the source data by
observed exposure in ascending order}, unexposed rows first, which is
the order the function imposes on its own copy (strategy B). The third
left the supplied order alone and \emph{corrected the function} instead,
building the exposure design matrix from the same sorted copy the
outcome design matrix already uses (strategy C). Pre-sorting and the
internal correction restore the same correspondence by different routes,
one changing the input and the other changing a single line of the
function. Because C differs from A only in the exposure-matrix ordering
substitution, the A--C contrast isolates the effect of that
substitution; the replicate-for-replicate agreement of B and C confirms
that pre-sorting restores the same correspondence.

\subsection{Cohorts and evaluation
measures}\label{cohorts-and-evaluation-measures}

We ran the three strategies on two source cohorts: a synthetic cohort of
4,000 subjects with six normally distributed covariates, simulated at
2,000 subjects, and the SUPPORT/Right Heart Catheterisation cohort of
5,735 complete cases with 32 covariates, simulated at
3,000\textsuperscript{11}. Each strategy used 200 replicates. The
injected exposure effect was null, so the true risk difference was zero.
This study used publicly available, de-identified SUPPORT/Right Heart
Catheterisation data and computer-generated data. No ethics approval was
required.

To judge whether the intended exposure--covariate relationship survived,
we applied the propensity score fitted in the source cohort to the
generated exposure and took the area under the receiver operating
characteristic curve (AUC). What this diagnoses is discrimination:
whether the generated exposure still carries the treatment--covariate
structure the source data encoded. The source score is deliberately not
the probability the implementation assigns to a given subject, since
that assignment is the thing under test. We also recorded achieved
exposure prevalence, covariate balance as the absolute standardised mean
difference (SMD), and bias in unadjusted and adjusted risk differences
(RD). Intervals around simulation means are 95\% \textbf{Monte Carlo
intervals}, quantifying uncertainty from the finite number of replicates
rather than uncertainty in a single study estimate\textsuperscript{1}.

\subsection{Additional experiments}\label{additional-experiments}

We used two additional experiments to assess the consequences of
misalignment. To test whether those consequences depend on the order
supplied, we ran the unmodified function on seven structured orders at
100 replicates each and on 500 independent random permutations at 10
replicates each, fixing the threshold for retained confounding before
the sweep. To test whether misalignment changes which estimators appear
biased, we estimated the risk difference five ways --- unadjusted,
g-computation and inverse-probability-of-treatment weighting, the latter
two each fitted once correctly and once with half the covariates
deliberately omitted --- under the unmodified and the
alignment-corrected generator, at 200 replicates each.

After correcting alignment, we used two additional experiments and
source inspection to assess problems that remained. To identify which
exposure drives the generated outcome, we repeated the
alignment-corrected simulation with an injected odds ratio of 2 and
fitted a model containing both the observed and the returned exposure,
at 200 replicates; those runs also let us reconstruct the contrast the
package reports as the truth and compare requested against achieved
exposure prevalence and outcome risk. We then ran the continuous-outcome
function \texttt{PlasmodeCont()} on the synthetic cohort at 100
replicates per strategy, to test whether it returns the generated
outcome and exposure under the right subject identifier, and inspected
the source of the survival function \texttt{PlasmodeSur()} without
running it. Designs and complete results for all of these are in Web
Appendices C--G.

\section{Results}\label{results}

\subsection{Row misalignment and its
effects}\label{row-misalignment-and-its-effects}

Source inspection showed that \texttt{PlasmodeBin()} builds the outcome
design matrix on a copy of the data sorted by observed exposure, but
builds the exposure design matrix on the rows as supplied, and then
applies indices drawn into the sorted copy to both (Table 1). A sampled
subject therefore receives another subject's exposure probability unless
the sort happens to leave their row in place. In both cohorts as
supplied, that is true of a single row in 4,000 and a single row in
5,735. The outcome compounds the problem rather than correcting it: it
is generated from the sampled subject's observed exposure, while the
separately drawn exposure is what the function returns.

In the supplied synthetic row order the unmodified generator nearly
erased the intended exposure--covariate association: the source
propensity score discriminated the generated exposure at an AUC of
0.503, against 0.841 once alignment was restored by either route (Table
2). Crude risk-difference bias moved with it, from +0.0051 to +0.3037.
The near-zero value is the one to be careful about, because in this
order it reflects the loss of the designed confounding rather than its
successful removal. The Right Heart Catheterisation cohort showed the
same loss and restoration of discrimination, and the same movement in
covariate balance, though its crude-bias magnitudes are far smaller.
Correcting the function reproduced pre-sorting replicate for replicate
on every recorded metric, confirming that pre-sorting restores the same
correspondence; the effect of the substitution itself is isolated by the
A--C contrast, which changes only that line. Balance moved in step: 19
of 32 Right Heart Catheterisation covariates exceeded a mean absolute
standardised mean difference of 0.1 after alignment, and none did under
the unmodified generator (Figure 1).

\subsection{Dependence on supplied row
order}\label{dependence-on-supplied-row-order}

Because misalignment permutes probabilities rather than removing them,
it does not always weaken confounding. What it does depends on the order
supplied. Achieved prevalence depends on the mean probabilities in the
positional blocks from which the sampler draws, not on whether
individual probabilities are assigned to the correct subjects. In the
order as supplied, those block means were close, and achieved prevalence
was 0.301 against a requested 0.300. Other orders gave different block
means and different prevalences, from 0.215 under propensity-ascending
order to 0.387 under propensity-descending order. A prevalence check can
therefore reveal misalignment, but it cannot rule it out. Across seven
structured orders, synthetic crude risk-difference bias spanned −0.0719
under exposure-descending order to +0.3032 under exposure-ascending
order, which is the aligned case, with +0.2858 under
propensity-ascending order and +0.0043 in the order as supplied. Across
500 random orders the mean AUC was 0.499, and none exceeded 0.55.

Misassignment and its consequence are separate questions: nearly all
subjects can receive another subject's probability even when that
probability stays close enough to the correct one to preserve most of
the intended confounding. Across the tested orders, misassignment is
near-total except under exposure-ascending, and what changes is how good
a substitute the wrong probability is. Under propensity-ascending order
97\% of subjects received another person's probability, yet it
correlated 0.94 with their own, and almost all the designed confounding
survived. Under exposure-descending order that correlation turned
negative, at −0.27, which is why the crude bias reversed sign. Web
Appendix C gives the full sensitivity results and the row mapping.

\subsection{Estimator comparisons}\label{estimator-comparisons}

We ran the five estimators on the synthetic cohort under both
generators. With alignment restored, the deliberately misspecified
estimators showed substantially more bias than the correctly specified
ones: g-computation carried a bias of +0.0005 and inverse probability
weighting +0.0085, while their misspecified counterparts carried
0.10--0.11 risk-difference bias. Under the unmodified generator the five
had similarly small mean biases. Across the four adjustment estimators
the spread of absolute estimated bias was 0.111 after correction against
0.002 without it. We compared bias only; variance and interval
calibration were not assessed. Web Appendix D gives all estimates.

\subsection{Faults remaining after
alignment}\label{faults-remaining-after-alignment}

Repairing the alignment does not make the generator correct. With
alignment corrected and an odds ratio of 2 injected, the injected effect
appeared on the \textbf{observed} exposure rather than the returned one:
log(2) = 0.693 showed up there as +0.696 (95\% Monte Carlo interval
0.676 to 0.716), while the returned exposure's conditional coefficient
was +0.017 (−0.006 to +0.041) and its adjusted risk difference +0.0025
(−0.0001 to +0.0051). The package separately reports a model
risk-difference contrast of +0.1105, omitting the intercept it used to
generate outcomes; the corresponding contrast at that intercept is
+0.0850. Those are two separable problems, and reinstating the intercept
fixes only the first: both contrasts describe changing the original
observed exposure in the outcome model rather than the simulated
exposure returned by the generator, so neither is its causal effect.
Both discrepancies are exactly zero at an injected odds ratio of 1. Web
Appendix F separates these estimands and documents a reconstruction
check.

A separate calibration problem remained after alignment correction: the
generator missed both the exposure prevalence and the outcome risk it
had been asked for. In the synthetic cohort achieved exposure prevalence
was 0.241 against a requested 0.300, and achieved outcome risk missed
\texttt{eventRate} the same way: 0.242 at an injected odds ratio of 1
and 0.224 at 2, against a requested 0.300. The source-cohort intercepts
are calibrated before sampling, which then draws within
observed-exposure strata, so the miss needs both a gap between the
target and the source prevalence and a separation between the calibrated
stratum probabilities; neither alone will produce it. Source prevalence
was 0.506 in the synthetic cohort, far from the requested 0.300. In the
Right Heart Catheterisation cohort, it was 0.381 against a requested
0.380, which explains the much smaller calibration miss. Sorting the
rows repairs alignment without removing this second fault; what it
changes is whether the fault is visible.

\subsection{Other package functions}\label{other-package-functions}

Evidence on the package's other two functions is uneven. An experiment
on \texttt{PlasmodeCont()} found the generated outcome and exposure
returned under another subject's identifier. \texttt{PlasmodeSur()}
builds no sorted copy and so never incurs the misalignment, but its
event times follow curves reflecting the observed exposure before a
replacement is drawn --- an observation from source inspection, not from
an experiment. Web Appendices E--G document the scope of both.

\section{Discussion}\label{discussion}

\subsection{Principal findings}\label{principal-findings}

In the branch we examined, the sampled subject's exposure probability is
usually read from another row, and the outcome is generated from the
exposure that subject actually had rather than from the one returned.
Neither departure raises an error or a warning, and nothing in the
returned data marks either one. The consequence falls on the
exposure--covariate association the generated data are supposed to
carry: in the synthetic cohort as supplied, the source propensity score
discriminated the generated exposure barely better than chance. Once
that association is gone, a crude estimate already sits near the truth
and adjustment has little left to remove. With alignment restored, the
comparison separated the deliberately misspecified from the correctly
specified estimators: misspecified adjustment carried 0.10--0.11
risk-difference bias, compared with +0.0005 for correctly specified
g-computation and +0.0085 for correctly specified inverse probability
weighting. Under the unmodified generator, the comparison no longer
separated correctly specified from deliberately misspecified adjustment
on estimated bias --- the dimension the benchmark exists to compare them
on.

The affected branch belongs to neither of the two designs Shaw et
al.~distinguish. Its outcomes are generated from the observed exposure,
which is what a sample-treatment design does, but the exposure it hands
back for analysis is a separate draw that never enters the outcome
model, so what the analyst receives is not sample-treatment data either.
That is also why repairing the row alignment restores one correspondence
without validating the generating mechanism as a whole: alignment
determines which probability a sampled subject's exposure is drawn from,
not which exposure the outcome is generated from. Shaw et al.~describe
design-level limitations of the sample-treatment framework; they do not
evaluate \texttt{Plasmode} software or row-order-dependent
implementation errors, and the defect reported here is specific to this
implementation of the generate-treatment design they
recommend\textsuperscript{2}.

\subsection{Practical checks}\label{practical-checks}

The results point to several checks for plasmode simulations. What made
the damage visible here was the exposure--covariate association measured
in the simulated cohorts rather than in the source data: the confounding
a method comparison depends on has to be present in the data the methods
are run on. Because the same call on the same data produced different
benchmarks under different input orders, the supplied row order should
be reported along with the package version. Where feasible, rerunning
the generator on a permuted copy of the input provides a direct
sensitivity check: the association, achieved prevalence and estimated
bias should not materially change. The comparison is distributional
rather than replicate by replicate, since a sound stochastic generator
attaches different random draws to different records once the rows move,
so identical output would be the wrong thing to require.

The three faults that survive alignment repair require separate checks.
Whether the exposure returned for analysis is the one the outcome was
generated from can be checked by fitting a model that contains both the
returned and the observed exposure, which is how the discrepancy
reported here became visible. The reported target should be
reconstructed from the coefficients and the calibrated intercept of the
model that generated the outcomes rather than accepted as the software
prints it. Requested and achieved marginal quantities --- exposure
prevalence and outcome risk --- should be compared directly, and that
comparison is not made redundant by the other two: the calibration miss
has a mechanism independent of row alignment and survives the repair, so
a generator can be correctly aligned and still miss the exposure
prevalence and the outcome risk it was asked for. In the supplied
synthetic order, achieved prevalence was essentially on target even
though 99.985\% of sampled subjects received another subject's
probability; the near-equality of the positional-block probability means
masked the calibration miss. Alignment and marginal calibration are
separate diagnostics, and either can be right while the other is wrong.

Each correspondence suggests a practical check. For row correspondence,
fit the source propensity model in the simulated cohort and compare its
discrimination with the source-cohort value. A large drop signals that
the intended exposure--covariate association was not preserved and
should prompt inspection of probability assignment. For intervention
correspondence, fit an outcome model containing both the returned and
observed exposure. An injected effect appearing on the observed rather
than returned exposure indicates that the wrong exposure generated the
outcome. For target correspondence, reconstruct the reported effect from
the coefficients and calibrated intercept of the generating model. As a
separate calibration check, compare requested and achieved marginal
quantities.

\subsection{Scope and implications}\label{scope-and-implications}

The evidence behind these recommendations covers one branch of one
function in depth and the rest of the package much more thinly. We
exercised the both-formula binary branch experimentally;
\texttt{PlasmodeCont()} had a single targeted experiment, and what we
say about \texttt{PlasmodeSur()} rests on source inspection alone.
Source inspection found no analogous mismatch in the formula-based
outcome-only branch, although the fitted-object outcome-only branch can
return an identifier that does not correspond to the covariates used to
generate the outcome; we did not test either branch experimentally, and
Web Appendix G gives the details. We did not conduct a systematic audit
of published applications, and the orders we swept were mostly
constructed or drawn at random rather than sampled from practice, so we
cannot say which orders analysts actually supply. The estimator panel
assessed bias, not variance or interval coverage. We did not repeat it
in the Right Heart Catheterisation cohort; its aligned crude bias was
+0.0132, compared with +0.3037 in the synthetic cohort.

The null effect we used for the main comparison conceals part of the
problem. At an injected odds ratio of 1 the outcome and reported-truth
discrepancies vanish, while the alignment defect and the calibration
misses remain visible; a simulation run only at the null would have left
two of the four faults described here visible and given no sign of the
other two. What mattered most for the method comparison, the lost
exposure--covariate association, was visible only in the simulated
cohorts and not in the arguments the generator was given. Checking a
generator therefore means checking the data it produced, not only the
model it was asked to produce them from.

\section*{Acknowledgements}

\textbf{Funding.} None declared.

\textbf{Conflicts of interest.} None declared.

\textbf{Use of artificial intelligence.} During this work, the author
used AI-based tools (large language models) to assist with the analysis
and simulation code, and text editing; the author verified all outputs
and takes full responsibility for the content.

\section*{Data Availability}

The SUPPORT/Right Heart Catheterisation data analysed in this study are
publicly distributed and are not redistributed here\textsuperscript{11}.
The synthetic cohort is generated from a recorded seed by the analysis
scripts. The analysis code, the archived replicate objects, the
generated outputs and the reproduction instructions are available in the
companion repository \texttt{plasmode-assess-code} at
\url{https://github.com/ehsanx/plasmode-assess-code}. The inspected
package source is the archived CRAN release of \texttt{Plasmode} 0.1.0,
fetched at run time and verified against a recorded SHA-256 rather than
redistributed.

\section*{Transparency and Reproducibility Statement}

Following the structure recommended by Wang and Pottegård (Am J
Epidemiol 2023), we report: (1) \emph{Protocol} --- no separately
registered protocol was prepared; the simulation design is reported in
the Methods and, following the ADEMP framework, in the Web Appendix. (2)
\emph{Preregistration} --- the study was not preregistered. (3)
\emph{Data access} --- the SUPPORT/Right Heart Catheterisation data are
publicly distributed\textsuperscript{11}; the synthetic cohort is
generated by the analysis scripts from a recorded seed. (4) \emph{Code
sharing} --- the analysis scripts, the archived replicate objects and
the generated tables and figures are in the companion repository
\texttt{plasmode-assess-code} at
\url{https://github.com/ehsanx/plasmode-assess-code}. The inspected
package source is the CRAN archive tarball, fetched at run time and
verified against its recorded SHA-256 before use rather than
redistributed. Analyses used R 4.5.1 and the archived CRAN release
\texttt{Plasmode} 0.1.0; the tarball digest, source provenance, the
one-line alignment substitution and the reproduction procedure are
documented in Web Appendices G--H and the \texttt{plasmode-assess-code}
repository. (5) \emph{Reporting guidelines} --- the simulation
experiments are reported following ADEMP\textsuperscript{1}.

\section*{References}

\protect\phantomsection\label{refs}

\clearpage

\section*{Tables}

\begin{table}[!h]
\centering
\caption{\label{tab:tab1-dataflow}Intended and actual data flow in the formula-based \texttt{PlasmodeBin()} generate-treatment branch. The package documentation states that when both models are supplied, the simulated outcome depends on the simulated exposure. Exposure prevalence and outcome risk are calibrated before sampling and are a separate matter from this sequence; see the Results. Source locations and the branch inventory are in Web Appendix G.}
\centering
\fontsize{8}{10}\selectfont
\begin{tabular}[t]{>{\raggedright\arraybackslash}p{2.4cm}>{\raggedright\arraybackslash}p{4.4cm}>{\raggedright\arraybackslash}p{6.0cm}}
\toprule
Step & Intended generate-treatment sequence & PlasmodeBin() 0.1.0, both-formula branch\\
\midrule
Assign exposure probability & Evaluate the exposure model on the sampled subject's covariates. & The exposure design matrix stays in the supplied row order, while the sampled indices refer to an exposure-ascending copy. Except at rows that ordering leaves fixed, the sampled subject receives another row's probability.\\
Draw exposure & Draw the subject's exposure from that probability. & Draw the returned exposure from the potentially misaligned probability.\\
Generate outcome & Generate the outcome from the exposure just drawn and the sampled subject's covariates. & Generate the outcome from the sampled subject's \textbf{observed exposure}, not the returned exposure.\\
Report target & Evaluate the contrast under the model that generated the outcome, including its calibrated intercept, for the exposure being simulated. & The reported contrast omits the outcome-calibration intercept. Restoring that intercept repairs that discrepancy, but the contrast still concerns the \textbf{observed exposure}, not the causal effect of the returned exposure.\\
\bottomrule
\end{tabular}
\end{table}

\begin{table}[!h]
\centering
\caption{\label{tab:tab2-alignment}Effect of row alignment on a null-effect plasmode benchmark. Each row summarises 200 replicates; values are means (95\% Monte Carlo intervals). Source and simulated sample sizes were 4,000 and 2,000 for synthetic data, and 5,735 and 3,000 for RHC. Requested exposure prevalence was 0.300 and 0.380, respectively; true RD was 0. The source-model AUC compares the source-cohort propensity score with generated exposure; 0.5 indicates no discrimination. In the source cohorts themselves that model gave AUCs of 0.837 (synthetic) and 0.756 (RHC); the aligned arms were close to those values. Bias is mean estimated RD minus true RD, in risk units. Strategy B sorts the source rows by observed exposure in ascending order, unexposed first, which is the order the function imposes on its own copy; B and C matched on every replicate. Adjusted RD bias and balance measures are in Web Tables A1–A2.}
\centering
\resizebox{\ifdim\width>\linewidth\linewidth\else\width\fi}{!}{
\fontsize{8}{10}\selectfont
\begin{tabular}[t]{llrrr}
\toprule
Cohort & Strategy & Exposure prevalence & Source-model AUC & Unadjusted RD bias\\
\midrule
Synthetic & A: unmodified, supplied order & 0.300 (0.299–0.301) & 0.503 (0.501–0.505) & +0.0051 (+0.0021 to +0.0081)\\
 & B: unmodified, exposure-ascending, unexposed first & 0.241 (0.240–0.242) & 0.841 (0.840–0.843) & +0.3037 (+0.3002 to +0.3072)\\
 & C: alignment-corrected, supplied order & 0.241 (0.240–0.242) & 0.841 (0.840–0.843) & +0.3037 (+0.3002 to +0.3072)\\
RHC & A: unmodified, supplied order & 0.380 (0.379–0.381) & 0.496 (0.494–0.497) & −0.0035 (−0.0061 to −0.0009)\\
 & B: unmodified, exposure-ascending, unexposed first & 0.381 (0.379–0.382) & 0.757 (0.756–0.758) & +0.0132 (+0.0106 to +0.0158)\\
\addlinespace
 & C: alignment-corrected, supplied order & 0.381 (0.379–0.382) & 0.757 (0.756–0.758) & +0.0132 (+0.0106 to +0.0158)\\
\bottomrule
\end{tabular}}
\end{table}

\clearpage

\section*{Figure}

\begin{figure}

{\centering \includegraphics[width=1\linewidth]{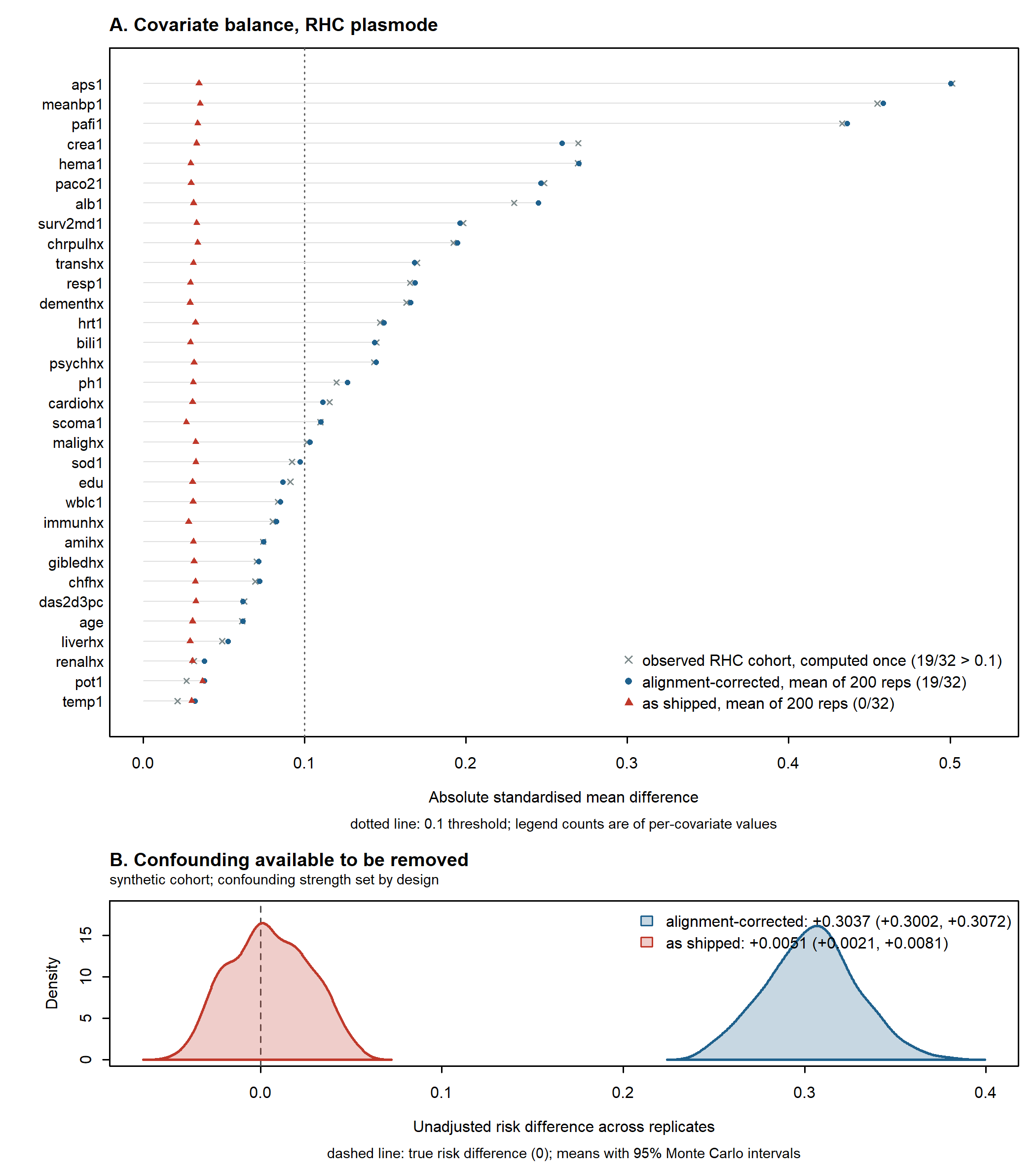} 

}

\caption{(A) Absolute SMD for each of 32 RHC covariates. The observed cohort contributes a single SMD per covariate, computed once on the source data, whereas each unmodified and alignment-corrected point is that covariate's mean over 200 simulated replicates. Values above 0.1 number 19 of 32 for the observed cohort, 19 of 32 after alignment correction and 0 of 32 unmodified. These are counts of per-covariate values, a different summary from the mean within-replicate percentage in Web Tables A1--A2, and the two should not be compared directly. (B) Distribution of the unadjusted RD across 200 synthetic replicates in the supplied row order. The alignment-corrected generator preserves the designed confounding, whereas the unmodified generator nearly removes it.}\label{fig:fig1-balance}
\end{figure}

Alt text: Two panels. Panel A is a dot plot of the absolute standardised
mean difference for each of 32 Right Heart Catheterisation covariates,
with one series for the observed cohort, one for the unmodified
generator and one for the alignment-corrected generator. The observed
and alignment-corrected series lie mostly above the 0.1 reference line,
while the unmodified series lies close to zero. Panel B overlays two
distributions of the unadjusted risk difference across 200 synthetic
replicates: the alignment-corrected distribution is centred well away
from zero, and the unmodified distribution is centred near zero.

\end{document}